\documentclass[runningheads]{llncs}

\usepackage{amssymb}
\usepackage{graphicx}
\usepackage{color}
\usepackage{url}
\newcommand{\keywords}[1]{\par\addvspace\baselineskip
\noindent\keywordname\enspace\ignorespaces#1}

\begin{document}

\mainmatter  % start of an individual contribution

\title{When Students Choose to Use Event-B in their Software Engineering Projects\thanks{This work was supported by grant ANR-13-INSE-0001 
(The IMPEX Project {\tt http://impex.gforge.inria.fr}) from the Agence Nationale de la Recherche (ANR).}}

\titlerunning{When Students Choose to Use Event-B}

%\author{J Paul Gibson\inst{1} \and  Idir A\¨{i}t-Sadoune\inst{2} \and Marc Pantel\inst{3}}
\author{J Paul Gibson\inst{1}}

%\authorrunning{Gibson, A\¨{i}t-Sadoune and Pantel}
\authorrunning{Gibsonl}

\institute{
SAMOVAR, T\'el\'ecom Sud Paris, CNRS, Universit\'e Paris Saclay,\\ 9 rue Charles Fourier, Evry Cedex,
91011 Paris, France\\
\url{paul.gibson@telecom-sudparis.eu}
 }

%\toctitle{Lecture Notes in Computer Science}
%\tocauthor{Authors' Instructions}
\maketitle

\begin{abstract}

Students often learn formal methods as part of a software engineering degree programme, without applying
these formal methods outside of the specific module(s) dedicated to this subject. In particular, software engineering
students often have to build a significant application/program/system in a substantial project
at the end of their programme (in order to
demonstrate the application of the things they have learned during the previous taught modules).
Our experience shows that the majority of students do not use formal methods in this project work.
We report on feedback from the minority of students who *did* choose to use formal methods in their projects,
and give examples of where this was a help and where it was a hindrance.
  
\end{abstract}

\section{Introduction}

This paper reports on the continuation of a sequence of  publications detailing the author's experience with
teaching formal methods. In 1998 \cite{GibsonMery98I} reports on the design and implementation of a
first (for the authors) formal methods course:
\begin{quote}
``Our approach to teaching formal methods tries to give an overall picture rather than concentrating on any one method, language or tool. We believe in letting the students discover the concepts and principles themselves, wherever possible''
\end{quote}

Two years later, our approach to teaching formal methods was  integrated  into a module dedicated to 
requirements engineering\cite{Gibson00}:
\begin{quote}
``Students are encouraged to question the need for formality --- each requirements engineering method is a compromise and the use of formal models needs to be placed within the context of the choices that a requirements engineer has to make''
\end{quote}

In \cite{Gibson08III} there is an overview of our approach to weaving formal methods throughout 
a software engineering programme, using problem based learning, and discussion of 
the impact of formal methods on the quality of the software that the students build:
\begin{quote}
``Anecdotal evidence suggests that the better students adopt formal engineering practices (like the specification of invariants) in projects on other courses which follow their work on the formal methods problems (without being told to do so). Furthermore, the software that these students produce is better than that produced by the other students. However, that should be no surprise as these are the better students!''
\end{quote}

In \cite{GibsonRaffy11}, we report on the design of a complete software engineering 
postgraduate degree programme, where rigour and formality are linked to modelling:
\begin{quote}
``All software engineering modules will be taught using a problem-based-learning (PBL) approach. Emphasis will be on rigour and formality, and mathematical modelling.''
\end{quote}
It was at this point in the development of our software engineering program that we decided to 
use Event-B\cite{Abrial10} and the Rodin tool\cite{AbrialBHHMV09}
 as our `default' formal method (even though we continued to
also use other methods).
The decision was based mainly on the positive feedback from various students regarding the RODIN tool, for
example:
\begin{quote}``{\em It was nice to have a formal methods IDE like Eclipse \dots  you can really experiment with
the models and the modelling process \ldots it makes  the maths more like programming \ldots its the first time
I understood the importance of invariants \ldots} etc.''
\end{quote}

This paper makes a novel contribution to this sequence of work/publications by reporting
on the analysis and feedback (from the students) that we have had since 2011.
We are not claiming that this is a scientific study; rather, we report on what we have observed,
what the students have stated during feedback interviews and after they have taken up employment
after graduating.

The remainder of this paper is structured as follows.
Section 2 provides a brief review of relevant related work in the teaching of formal methods. Section 3 motivates the need for the type of study being reported in this paper.
Section 4 provides information concerning the students who have participated in this study (through
the feedback that they have provided). Section 5 is the main contribution of the paper, where we review
the key observations and lessons to be learned. In section 6, we conclude with some
recommendations for teachers of formal methods.

\section{Related Work - teaching formal methods}

In this section we report on previous work that has had the most influence on our own
approach to teaching formal methods. It is not intended as a comprehensive 
review of  the history and state-of-the-art.

It is important to note our work is concerned with teaching formal methods to (software) engineering students
and not to computer science students\cite{Parnas99}. Curriculum design for software engineering students
requires making complex trade-offs between the teaching of theory and practice\cite{Garlan92,Garlan95}.
One of the first books dedicated to the subject of teaching formal method\cite{DeanHinchey96} identifies the
role that teachers play in improving the transfer of formal methods technology to industry, through their students.
At the turn of the century, a sizeable community of formal methods teachers had grown and 
started to organise their own workshops concerned with establishing 
formal methods as a key part of the SoftWare Engineering Body Of Knowledge (SWEBOK)\cite{AlmstrumDGHS00}.
The need for a ``{\em A Different Software Engineering
Text Book}''  was identified by Bjorner \cite{Bjorner01}.

The need for scientific evidence supporting the importance of teaching formal methods to software engineers
was highlighted by Henderson \cite{Henderson03}:
\begin{quote}``Evidence supporting the importance of mathematics in software engineering practice is sparse. This 
naturally leads to claims that software practitioners don’t need to learn or use mathematics. Surveys of current practices reflect reality; many software engineers have not been taught to use discrete mathematics and logic as effective tools. Education is the key to ensuring future software engineers are able to use mathematics and logic as power{ful} tools for reasoning and thinking.''
\end{quote}

Before Event-B there was B\cite{Abrial96a}. Teaching formal methods using B is reported in a number of
papers, including:
\cite{LeuschelSBL08,Habrias08I,mery2008,GuyomardAHJS09}

\section{Motivation: technology transfer and best practice}

The important role of students in the transfer of software engineering
technology to industry was illustrated by \cite{HallinanGibson05},
where the technology in question was UML. Parnas has argued that technology transfer of formal methods will
fail because ``{\em We can’t sell methods that we don’t use ourselves.}''\cite{Parnas98II}. 
Our view is a reworking of the phrase from Parnas -- our students can't sell formal methods if they
don't choose to use them themselves.

Consequently, we wished to observe whether our students choose to use formal methods when
working on assessed projects that required the development of software. 

\section{The Educational Context for our Observations}

The MSc program was a 2 year program which ran between  2010 and  2014.
The student intake was global from 4 continents --- Europe, Africa, Asia and the Americas.
Entry to the program was highly selective, with an acceptance rate of between 10 and 20 percent.
 Subsequently,
 the number of students in each year was relatively small
with an average of 8 per year.
As a consequence of the small number of students, our analysis is not based on a scientific (statistically significant) study.
Instead, we report on the feedback from students gathered through questionnaires, interviews and 
informal communication.

\section{Observations and Lessons}

We structure the observations based on whether the feedback was concerned with project work,
placement work or work since graduation.

\subsection{Use of formal methods in project work}

At the end of the program, the students are expected to work on a significant software engineering
project (3-person months per student). They can choose to work in teams or individually.
They are free to use whatever techniques/tools/languages/processes that they wish, but they must justify
their choice based on the exact nature of the project on which they were working.

After seeing formal methods throughout the program, as well as having a module dedicated to 
teaching them Event-B and Rodin, we were hoping that the majority of students would 
write formal specifications in order to model key requirements and/or design issues.

Over the 4 year period, only 3 projects from a total of 14 incorporated significant models in Event-B.
These 3 projects were ranked (over the 4 years) in places 1, 3 and 13.
The 2 `top' projects were submitted by the best students (based on performance on all modules).
They chose to write Event-B models because:
`` {\em \ldots we wanted to get a better understanding of the rules of the game that we were
developing.''}, and ``{\em \ldots the application was safety critical and we wanted to be sure that
the design of the communication protocol was correct.''}. For the highest marked project,
the team produced a poker game, modelled the rules formally and verified that the operator for ranking
hands was based on a transitive relationship. During their development, the RODIN tool helped
them identify `bugs' in their models concerned with misunderstanding of different types of hand.
The second project using formal methods developed an android application for use by
emergency services when arriving at the scene of an accident. A main issue was how 
data could be communicated to/from the hospital as effectively as possible.
They designed a protocol for the communication but worried that it could lead to 
deadlocks in the interface. They successfully modelled the protocol in Event-B but were
unable to express (or consequently prove) the required property.
The project ranked 13 was submitted by a group who chose to use formal methods because
they thought that: `` {\em \ldots that was what the teacher was looking for.''}. 
They worked on a parallel implementation of a genetic algorithm for pattern recognition.
Unfortunately, their lack of experience and ability in formal methods meant that they never
finished the specification phase of development, and when they started design and
code they were very behind schedule.

Students from projects that chose not to use formal methods were interviewed after
they received their evaluations. Two of the groups regretted not writing a formal specification because
they had significant problems arising from the team members having inconsistent understanding
of their requirements. All groups reported choosing not to use them because they
didn't feel that they needed them, and that they wanted to use more agile development
approaches (which they felt were not suited to formal methods).

\subsection{Use of formal methods during placement}

Through analysis of the student placement reports, and through
their presentations, we were able to evaluate the degree of use of formal
methods by the students during their placements. We classified the use at 4 different levels:
\begin{enumerate}
\item Using formal methods was a critical requirement of the placement (2 students)
\item The student was required to use  formal concepts, such as invariants in code, during their  placement but there was no dedicated
        formal methods tool (6 students)
 \item The student was not required to use formal methods, but they were able to use them in their own work. (1 student)
 \item The student was not required to use formal methods, and  did not use them (20+ students)
 \end{enumerate}
 
 It is, perhaps, not surprising that so few students used formal methods during their placements.
 The 2 students who were obliged to use them had been placed in research and development 
 environments (in education and in industry) where formal methods tools were being developed.
 The 6 students who were required to use formal concepts were working in safety-critical
 domains such as the aerospace and health sectors. The one student who chose to try and
 use formal methods, even though they were not required, reported: ``{\em a certain frustration that my 
 co-workers found it amusing that I would wish to use mathematical models}''.

\subsection{Use of formal methods after graduating}

A significant minority of students(8 in total) stay in regular contact with us after graduating. 
None of them are working in an environment which uses formal methods.
A handful of them believe that the quality of their work would be improved
through the use of formal methods.

\section{Conclusions: recommendations for teachers}

Although our report is based on a small number of observations, it is worrying that Parnas appears to 
be (at least partially) correct
 when he stated that we will not be able to transfer formal methods technology from academia to
industry.

It is not the teachers' role to force their students to use formal methods. Successful teaching
of formal methods will  motivate students to use them because they believe in them.
We, as teachers, need to better monitor students during the whole of their academic careers (and after)
to measure the use of formal methods, together with the impact of their use on the quality of
software being developed. We also need to better support students who wish to 
introduce formal methods technologies in their workplace.

\bibliographystyle{splncs}

\bibliography{Event-B-2016}

\end{document}